| RESEARCH ARTICLE

# Impacts of Business Architecture in the Context of Digital Transformation: An Empirical Study Using PLS-SEM Approach

**Dennis O'Higgins**

*Consulting Capability Leader, Strategy & Operations, Slalom Consulting, Bloc 17 Marble Street, Manchester, United Kingdom; HPA, Salford Business School, The University of Salford, Maxwell Building, Salford, United Kingdom; DBA Scholar, Swiss School of Business & Management, Geneva Business Center, Avenue Des Morgines 12, Geneva, Switzerland*

**Corresponding Author:** Dennis O'Higgins, **E-mail**: d.j.ohiggins@salford.ac.uk

| ABSTRACT

Despite the critical importance of Digital Transformation, up to 95% of initiatives fail to deliver expected business benefits. This paper explores the role of Business Architecture practices in enhancing digital transformation success. Using an adapted Balanced Scorecard approach and a Structural Equation Model (SEM), we analysed survey responses from 129 industry practitioners using a Partial Least Squares (PLS) approach. Our findings indicate that effective business architecture practices significantly improve business alignment, efficiency, service delivery, and strategic outcomes, leading to successful digital transformation. The study also validates factors proposed by AL-Malaise AL-Ghamdi (2017) in the context of digital transformation. The paper presents an adapted conceptual model addressing discriminant validity issues in previous models and benefiting from the robustness of the Balanced Scorecard approach. The study concludes by highlighting the essential role of business architecture in driving digital transformation success.

| KEYWORDS

Business Architecture, Digital Transformation, Strategic Alignment, Technology Strategy, Business Impact, Partial Least Squares, Structural Equation Modelling.

## 1. Introduction

Digital transformation (DT) has become a crucial aspect of almost every organizational strategy, as businesses seek to adapt to the shift towards Industry 4.0 and the rapidly growing market demand for digital solutions. In a bid to keep pace with demand, organisations are also seeking to make sense of the potential new commercial opportunities such solutions can present and subsequently capitalize on those opportunities (Schallmo et al., 2017). By 2026, business leaders are expected to invest $3.4 trillion USD in their DT programmes, which represents a 16.3% increase, when measured on a five-year compound annual growth rate (CAGR) (Shirer, 2022). Despite this investment, most digital transformations do not achieve the benefits anticipated in their initial business cases, with estimates of the failure rate ranging from 70% to 95% (Bonnet, 2022). Underpinning the difference between estimates is the degree of conservatism when categorizing failure – ranging from clear, abject failure across multiple measures to minor discrepancies against only some minor metrics. Accordingly, there is an increasing interest in improving the outcome of DT programmes (Bonnet, 2022; Schallmo et al., 2017; Tangi et al., 2020; Vial, 2019).

Business architecture, which refers to the design and structure of an organization's processes, capabilities, and systems (Hendrickx et al., 2011), is beginning to emerge as a key component of DT. Effective business architecture is critical for organizations to successfully implement digital transformation initiatives, as it enables them to align their strategies, processes, and systems with their digital objectives (AL-Malaise AL-Ghamdi, 2017; Amit & Zott, 2015; Vial, 2019). Nevertheless, there is a significant gap in our understanding of its impacts (AL-Malaise AL-Ghamdi, 2017; Gong et al., 2020; Vial, 2019). As such, this study seeks to investigate the impacts of business architecture in the context of digital transformation. Specifically, we aim to understand the relationship between business architecture and key outcomes to gain insights into the role of business architecture in driving successful DT.

This study's findings provide important implications for both academics and practitioners. From an academic perspective, this study contributes to the theoretical understanding of business architecture and its role in DT. The study's results extend our knowledge of how organizations can design effective business architectures in the context of DT. Equally from a practical perspective, this study's findings are of great interest to organizations undergoing or planning DT. The results will provide valuable insights into the importance of business architecture in facilitating successful DT and offer guidance on how to design and implement effective business architectures. Ultimately, this study advances our understanding of the impacts of business architecture in the context of DT and provides practical recommendations to organizations seeking to undertake this critical transformational process.

## 2. Literature Review

Digital transformation is inherently disruptive for organisations (Karimi & Walter, 2015; Naimi-Sadigh et al., 2022; Vial, 2019) and a review of the extant literature suggests a major root cause of challenges for DT programs is a common, but pervasive misconception about the focus and objectives of DT. Specifically, the notion that the transformation is wholly or primarily focused on the underlying technology itself and success is thus defined or bounded by the same (Agrawal et al., 2020; Kane, 2019; Tangi et al., 2020; Vial, 2019). Whereas the literature reveals technology alone is seldom the substantive cause of "failure" (Zhu et al., 2021). Instead, DT is complex and multi-faceted, with other factors, beyond the technology, most often being the source of friction and hindering the realisation of the benefits case. Following a 4 year-long study with over 16,000 participants Kane (2019), concluded that whilst technology is clearly an important enabler of DT, it's success and the resulting digital maturity of the organisation is determined by other factors, which Vial (2019) suggests can be broadly categorized as either structural factors or organisational barriers. Structural factors can be considered as the fabric of the organisation and include issues such as: Organizational Structure (Maedche, 2016), Culture (Hartl & Hess, 2017), Leadership (Horlacher et al., 2016), Roles, Responsibilities and Skills (Dremel et al., 2017; Hess et al., 2016). Whereas barriers, are the typical impediments that prevent progress against the structural factors and include: Inertia (Svahn et al., 2017), Resistance (Fitzgerald et al., 2014), Risk Tolerance (Fehér & Varga, 2017), Collaboration and Experimentation (Kane, 2019). Vial (2019) and Kane (2019) also highlight those organizations reporting greater digital transformation maturity also exhibit more strategic focus through a relatively greater mastery of dynamic capabilities within their organization. Dynamic capabilities were originally defined by Teece et al. (1997) as "the firm's ability to integrate, build, and reconfigure internal and external competences to address rapidly changing environments". This builds on Kane's (2019) research which found that a critical factor for DT outcomes is the ability to be flexible and adapt the organisation regularly and thus keep pace with the rate of technological change over technology excellence. Subsequently dynamic capabilities have been categorized into three broad categories: sensing, seizing and transforming (Teece, 2007) referring to the organizations ability to identify and anticipate change, to mobilize resource to capture value and to continually (re-)align and (re-)deploy assets respectively.

### 2.1 The developing role of Business Architecture in Digital Transformation

As organizations have embarked on digital transformation, there has been a growing recognition of the importance of business architecture in driving successful outcomes (Gong et al., 2020; Vial, 2019; Zhu et al., 2021). Business architecture can help organizations understand their current state, identify gaps, and design a future state that aligns with their digital transformation objectives (Amit & Zott, 2015). Business architecture plays a vital role in the digital transformation of an organization by helping to align business and IT strategies, processes, and systems (AL-Malaise AL-Ghamdi, 2017). In the context of DT, business architecture can provide an effective framework for aligning an organization's strategic objectives with its digital transformation initiatives. Effective business architecture enables organizations to design, implement and monitor their digital transformation initiatives in a structured manner, ensuring that they deliver the intended benefits while avoiding the risks and challenges associated with the transformation process (Bodine & Hilty, 2009; Hendrickx et al., 2011; Tortora et al., 2021). Business architecture also facilitates cross-functional collaboration, thereby enhancing communication and reducing the risks associated with siloed working. Business architecture can also enable an organization to design and implement a flexible and adaptable operating model that can respond effectively to changing market conditions, emerging technologies, and new business models (Amit & Zott, 2015; Berman, 2012; Schallmo et al., 2017). By aligning the various elements of the organization with its digital objectives, business architecture can help ensure that the organization remains agile, innovative, and competitive. The role of business architecture in DT is to provide a structured approach for aligning an organization's business and IT strategies, processes, and systems with its digital transformation objectives. Effective business architecture can help organizations deliver the intended benefits of their digital transformation initiatives while avoiding the risks and challenges associated with the transformation process.

*2.2 The Initial Conceptual Model for Measuring Business Architecture Impact and its Limitations*
To measure the impact of business architecture on digital transformation, our initial approach adopted the conceptual model proposed by (AL-Malaise AL-Ghamdi, 2017). This model comprises three main components: the operating model, the IT model, and the strategic model. The operating model focuses on the daily operations and processes of an organization, the IT model emphasizes the role of technology and information systems, and the strategic model deals with the alignment of business strategies and objectives.

We initially considered this model to be a promising framework due to its comprehensive approach in encompassing operational, technological, and strategic aspects of an organization undergoing digital transformation. However, we encountered issues during the empirical testing of this model. The critical issue arose with the with discriminant validity. We observed that the distinctions between the operating model, IT model, and strategic model were not as clear as hypothesized, and these components were not distinctly measurable. This limitation raised concerns regarding the reliability and applicability of the model for accurately measuring the impact of business architecture in the context of digital transformation. This necessitated the exploration of alternative frameworks using the data collected that could effectively address these limitations

*2.3 The Theoretical Framework and Adapted Conceptual Model for Measuring Business Architecture Impact*
Following the limitations encountered with the AL-Malaise AL-Ghamdi (2017) model, we reviewed the underlying theoretical framework to explore alternative approaches for the conceptualization of impact through the application of business architecture practices. The Balanced Scorecard (BSC) method proposed by (Kaplan & Norton, 1992), was identified as a suitably robust alternative for developing an adapted conceptual model. The BSC is a widely used performance measurement and management tool that enables organizations to align their strategic objectives with their operational activities (Kaplan & Norton, 2005). The BSC measures an organization's performance across four perspectives: financial, customer, internal processes, and learning and growth. Moreover, the BSC is known for its ability to measure organizational performance across multiple dimensions (Hasan & Chyi, 2017).

In the adapted model, we aligned the dimensions of the BSC with the context of digital transformation. In the context of DT, we adapted the BSC to measure the impact of business architecture on digital transformation across four dimensions: business alignment, efficiency, service delivery, and strategic outcomes. Business alignment measures the extent to which business and IT strategies are aligned. Efficiency measures the cost and cycle time reduction achieved through digital transformation. Service delivery measures the improvement in customer service and quality. Finally, strategic outcomes measure the competitive advantage achieved through digital transformation. We selected these dimensions based on the conceptual model proposed by AL-Malaise AL-Ghamdi (2017) and the literature on the impact of business architecture on DT. We believe that these dimensions capture the key benefits of business architecture in the context of DT and provide a comprehensive framework for measuring its impact.

Our adapted conceptual model consists of four latent variables, namely business alignment, efficiency, service delivery, and strategic outcomes, which is gauged by a total of 15 indicators. Figure 1 illustrates the proposed model. This adapted model addresses the discriminant validity issue encountered in the AL-Malaise AL-Ghamdi (2017) model by providing clearer distinctions between the dimensions. It also benefits from the established robustness of the Balanced Scorecard approach.

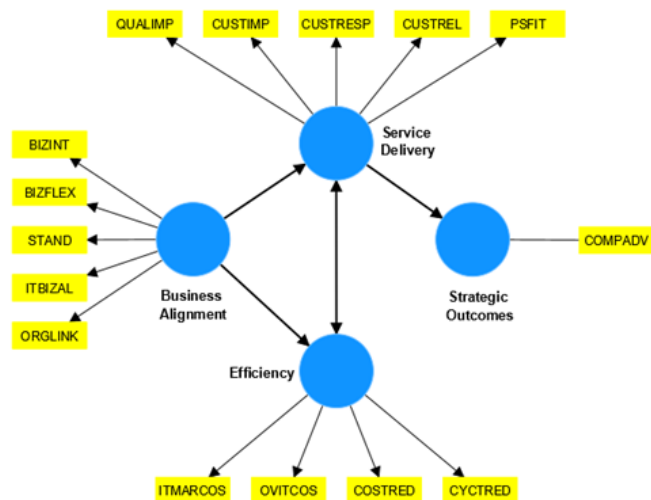

**Figure 1: Adapted Conceptual Model for Measuring Business Architecture Impact**

This adapted conceptual model is aimed at measuring the extent to which business architecture aligns with digital transformation goals, enhances operational efficiency, improves service delivery, and facilitates achieving strategic objectives. The 15 indicators capture the key dimensions of business architecture that can have a significant impact on digital transformation success. The model serves as a useful tool for organizations to assess the effectiveness of their business architecture in driving digital transformation and to identify areas for improvement.

*2.4 Research Questions and Hypotheses*
In the context of digital transformation, this study aims to explore the impact of business architecture on organizational performance. Based on the literature review, theoretical background and our conceptual model, the following research questions and hypotheses have been formulated:

**Table 1: Research questions and corresponding hypotheses**

| Research Questions | Corresponding Hypothesis |
| --- | --- |
| RQ1: How does Business Alignment influence Service Delivery in the context of digital transformation? | H1: Business Alignment has a positive impact on Service Delivery. |
| RQ2: How does Business Alignment impact Efficiency in organizations undergoing digital transformation? | H2: Business Alignment has a positive impact on Efficiency. |
| RQ3: What is the role of Efficiency in enhancing Service Delivery during digital transformation initiatives? | H3: Efficiency has a positive impact on Service Delivery. |
| RQ4: How does improved Service Delivery contribute to achieving Strategic Outcomes in the context of digital transformation? | H4: Service Delivery has a positive impact on Strategic Outcomes. |

Hypothesis 1: This hypothesis posits that organizations with better Business Alignment in their digital transformation initiatives will experience improved Service Delivery, encompassing aspects such as quality improvement, customer service enhancement, customer responsiveness, customer relations, and product/service fit.

Hypothesis 2: Our second hypothesis suggests that enhanced Business Alignment within the organization during digital transformation efforts will lead to increased Efficiency, as reflected in reduced marginal and overall IT costs, and improved cost reduction and cycle time reduction.

Hypothesis 3: This hypothesis posits that organizations with higher Efficiency during digital transformation will exhibit improved Service Delivery, as the effective utilization of resources and streamlined processes contribute to better customer service and responsiveness.

Hypothesis 4: Our fourth hypothesis proposes that improved Service Delivery resulting from digital transformation initiatives will contribute to the achievement of Strategic Outcomes measured by reported competitive advantage.

## 3. Methodology
We tested these hypotheses using a quantitative research design with data collected via a digital survey and analysed using a partial least squares structural equation modelling (PLS-SEM) approach.

*3.1 Development of the Research Instrument*
As a starting point, we took the conceptual model proposed by AL-Malaise AL-Ghamdi (2017) and developed a questionnaire to operationalize this by measuring each of the 15 benefits identified in their model. We deployed a 5-point Likert scale (strongly disagree, disagree, neutral, agree and strongly agree) to assess each. Ordinarily, researchers seek to remove the middle or neutral option in Likert scales to force respondents to make a choice (Croasmun & Ostrom, 2011). However, Randall & Fernandes (1991) highlighted this can lead to inadvertent response bias and this is especially problematic where a topic is especially ambiguous or obscure (Johns, 2005). This is relevant to Business Architecture since its deployment can vary greatly between organizations and in many cases, it is not practised at all. By providing a neutral option, we enable respondents to show they have not formed an opinion, which in the context of our research, is a relevant finding in itself (Chyung et al., 2017). Additionally, we collected biographical information from respondents, including if the respondent worked in business architecture, their industry, the number of employees in their company and the annual revenue of their company. Respondents were also asked to select the number of years the firm has been practicing business architecture, the relative maturity of those business architecture practices and if business architecture was considered structurally part of the IT organisation in their company. The questionnaire was reviewed by an academic expert in quantitative analysis and a practitioner with expertise in business architecture and academic research.

Subsequently, some minor adjustments were made so the statements were unambiguous and capable of being easily understood by respondents.

### *3.2 Sampling and Data Collection*

The survey was distributed online using SurveyMonkey via a series of Business Architecture, Technology and Leadership related interest groups on the LinkedIn platform (table 2), through the Business Architecture Guild's member community and via direct messages to members of the researchers' professional LinkedIn network. Survey responses were collected between 19 June 2022 to 19 August 2022. Responses were screened for completeness and any respondents who did not provide a response for every question were excluded from further analysis. 169 respondents engaged with the digital questionnaire; 40 responses (23.7%) were excluded during completeness screening, and the remaining 129 valid responses (76.3%) were analysed further. We used the statistical algorithm developed by (Westland, 2010) to compute the absolute minimum sample size. This was determined to be 116 cases, based on 4 latent variables and 15 indicator variables with a statistical power of 0.8 and significance of 0.05. Accordingly, our sample size of 129 can be determined to meet the sampling adequacy requirements (Westland, 2010).

Table 2 - LinkedIn Groups used to disseminate the research questionnaire.

| LinkedIn Group |
| --- |
| Business Architecture Perspectives |
| Business Architecture Community |
| Enterprise Architecture Forum |
| Association for Project Management Discussion Forum |
| ISPIM Innovation Community Group |
| ISPIM Digital Disruption and Transformation Special Interest Group |
| Chief Information Officer (CIO) Network |
| CEO Network |
| Chief Financial Officer Network |
| Organization Design Forum |
| Change Management UK |
| Digital Strategy & Transformation |

### 4. Results and Discussion

We initially assessed the data using StataSE version 17.0 to establish the descriptive statistics and reliability. We intended to use Structural Equation Modelling (SEM) for our analysis and so assessed the data for normality since a central requirement of conventional SEM is for the data to be normally distributed (Blanthorne et al., 2006). In our data, Skewness ranged from -0.955 to -0.091 and kurtosis from 2.04 to 3.88. As several of our values for kurtosis exceeded 3, our data was considered to violate the criteria for normality set out by (Kline, 2016). Therefore, we adopted a Partial Least Squares (PLS) based SEM for the analysis. PLS was selected for its ability to model under conditions of non-normality and relatively smaller sample sizes (Hair et al., 2013). Ali et al. (2016), highlight PLS to be well-established as a technique within the field of Management research and is also widely adopted in other fields for the purposes of structural modelling and path analysis. We conducted our PLS analysis using SmartPLS 4.0 software and followed the two-stage analytical procedure espoused by Hair et al. (2013). We first considered the measurement model to establish the reliability and validity of the measures themselves before analysing the structural model. In our analysis, to test the significance of the path co-efficients and loadings we employed the bootstrapping method in SmartPLS 4.0 and used 5000 resamples (Hair et al., 2013). We have assessed and presented our results in accordance with the "guidelines when using PLS-SEM" developed by Hair et al. (2019).

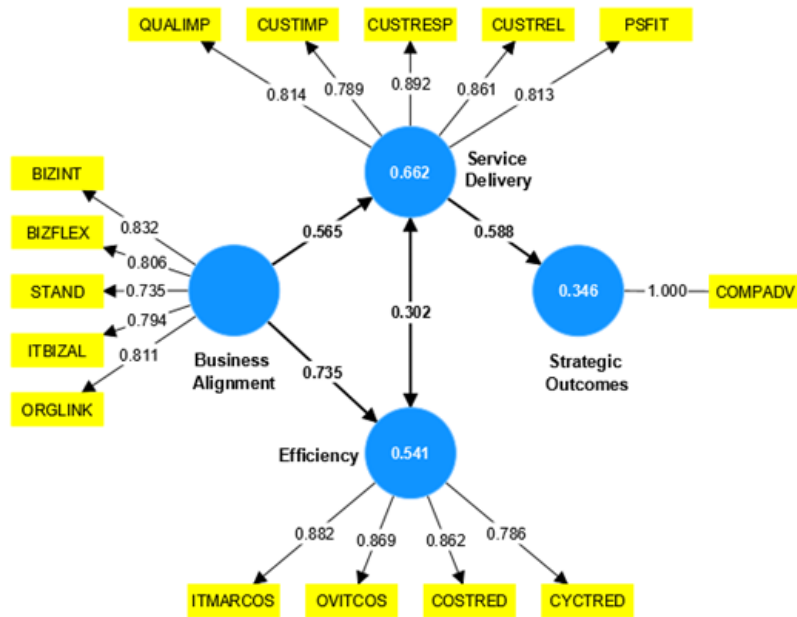

**Figure 2: Results of empirical assessment using the structural model**

*4.1 Measurement Model*

The measurement model is assessed by examining the reliability, convergent validity, and discriminant validity of the constructs. The loadings, Average Variance Extracted (AVE), and Composite Reliability (CR) values for each construct are presented in Table 3a.

All loadings are above the 0.7 threshold, indicating acceptable item reliability (Hair et al., 2019). AVE values are above 0.5, and CR values are above 0.7, which demonstrates convergent validity (Fornell & Larcker, 1981).

Discriminant validity is assessed using Fornell-Larcker criterion (Table 3b), Heterotrait-Monotrait (HTMT) ratio (Table 3c), and cross-loadings (Table 3d). According to the Fornell-Larcker criterion, the square root of AVE values (bolded diagonal values) should be higher than the off-diagonal correlations (Fornell & Larcker, 1981). Table 3b shows that this condition is met for all constructs. Additionally, Table 3c presents the HTMT values. Although the strict criterion for HTMT is 0.85, some researchers argue that the threshold should be set at 0.90 as an absolute value (Henseler et al., 2015). In our study, all HTMT values are below 0.90, which suggests that discriminant validity is supported. While one of the HTMT values (Business Alignment – Service Delivery) are slightly above the strict 0.85 criterion, it is essential to note that it is still below the absolute 0.90 threshold. The slight deviation from the strict criterion might be due to the high correlations between the constructs, which is not uncommon in practice. However, further analysis might be needed to ensure the distinctiveness of these constructs.

Table 3d presents the cross-loadings, which provide additional support for discriminant validity. The cross-loadings indicate that each item loads higher on its respective construct than on any other construct (Hair et al., 2021) (Hair et al., 2017). Collinearity was examined using the Variance Inflation Factor (VIF) values (Table 3e), which were below the recommended threshold of 5, indicating no multicollinearity issue among the constructs (Hair et al., 2021).

**Table 3a: Model Loadings and Assessment**

| Construct | Measurement Items | Loadings | AVE | CR |
|---|---|---|---|---|
| Business Alignment | BIZINT: Business Integration | 0.832 | | |
| | BIZFLEX: Business Flexibility | 0.806 | | |
| | STAND: Standardization | 0.735 | 0.634 | 0.858 |
| | ITBIZAL: Alignment of IT and business strategy | 0.794 | | |
| | ORGLINK: Organizational links | 0.811 | | |
| Efficiency | ITMARCOS: Marginal cost of IT | 0.882 | | |
| | OVITCOS: Overall cost of IT | 0.869 | 0.723 | 0.872 |
| | COSTRED: Cost reduction | 0.862 | | |
| | CYCTRED: Cycle time reduction | 0.786 | | |
| Service Delivery | QUALIMP: Quality improvement | 0.814 | | |
| | CUSTIMP: Customer service improvement | 0.789 | | |
| | CUSTRESP: Customer responsiveness | 0.892 | 0.697 | 0.893 |
| | CUSTREL: Customer relations | 0.861 | | |
| | PSFIT: Product/service fit | 0.813 | | |
| Strategic Outcomes | COMPADV: Competitive advantage | 1.000 | | |

**Table 3b: Discriminant Validity**

| Construct | Business Alignment | Efficiency | Service Delivery | Strategic Outcomes |
|---|---|---|---|---|
| Business Alignment | **0.796** | | | |
| Efficiency | 0.735 | **0.851** | | |
| Service Delivery | 0.787 | 0.718 | **0.835** | |
| Strategic Outcomes | 0.525 | 0.4 | 0.588 | **1.000** |

Values on the diagonal (**bolded**) are square root of the AVE while the off-diagonals are correlations

**Table 3c: Heterotrait-monotrait (HTMT)**

| Construct | Business Alignment | Efficiency | Service Delivery | Strategic Outcomes |
|---|---|---|---|---|
| Business Alignment | | | | |
| Efficiency | 0.842 | | | |
| Service Delivery | 0.895 | 0.810 | | |
| Strategic Outcomes | 0.567 | 0.426 | 0.623 | |

**Table 3d: Cross-loadings**

| Indicator | Business Alignment | Efficiency | Service Delivery | Strategic Outcomes |
|---|---|---|---|---|
| BIZFLEX | 0.806 | 0.657 | 0.622 | 0.393 |
| BIZINT | 0.832 | 0.524 | 0.618 | 0.505 |
| COMPADV | 0.525 | 0.4 | 0.588 | 1 |
| COSTRED | 0.649 | 0.862 | 0.637 | 0.338 |
| CUSTIMP | 0.602 | 0.552 | 0.789 | 0.503 |
| CUSTREL | 0.6 | 0.592 | 0.861 | 0.472 |
| CUSTRESP | 0.723 | 0.678 | 0.892 | 0.519 |
| CYCTRED | 0.64 | 0.786 | 0.64 | 0.415 |
| ITBIZAL | 0.794 | 0.512 | 0.547 | 0.347 |
| ITMARCOS | 0.591 | 0.882 | 0.579 | 0.277 |
| ORGLINK | 0.811 | 0.64 | 0.698 | 0.396 |
| OVITCOS | 0.614 | 0.869 | 0.576 | 0.32 |
| PSFIT | 0.655 | 0.581 | 0.813 | 0.483 |
| QUALIMP | 0.696 | 0.585 | 0.814 | 0.475 |
| STAND | 0.735 | 0.57 | 0.631 | 0.447 |

**Table 3e: Collinearity statistics (VIF)**

| Construct | Business Alignment | Efficiency | Service Delivery | Strategic Outcomes |
|---|---|---|---|---|
| Business Alignment | | 1.000 | 2.177 | |
| Efficiency | | | 2.177 | |
| Service Delivery | | | | 1.000 |
| Strategic Outcomes | | | | |

*4.2 Structural Model*
The structural model is assessed by examining the path coefficients, t-values, R-squared, Adjusted R-squared, effect sizes ($f^2$), and predictive relevance $Q^2$ (PLSPredict) values.

The path coefficients and t-values are presented in Table 3f. All four hypotheses are supported, as the t-values exceed the critical value of 1.96 (P<0.05). The R-squared ($R^2$) and Adjusted R-squared (Adj. $R^2$) values in Table 3g indicate the proportion of explained variance in the dependent constructs. The values suggest a moderate-to-strong level of explained variance in Efficiency ($R^2$ = 0.541), Service Delivery ($R^2$ = 0.662), and Strategic Outcomes ($R^2$ = 0.346) (Hair et al., 2019). Effect sizes ($f^2$) in Table 3h demonstrate the relative impact of each predictor on the dependent constructs. An $f^2$ value of 0.02, 0.15, and 0.35 represents a small, medium, and large effect, respectively (Cohen, 1988).

The $f^2$ values in Table 3h suggest that Business Alignment has a large effect on Efficiency ($f^2$ = 1.177) and Service Delivery ($f^2$ = 0.434), while Efficiency has a small-to-medium effect on Service Delivery ($f^2$ = 0.124). Furthermore, Service Delivery has a medium-to-large effect on Strategic Outcomes ($f^2$ = 0.529). The predictive relevance of the model is assessed using the $Q^2$ values obtained from PLSPredict (Table 2i). $Q^2$ values greater than zero indicate that the model has predictive relevance (Hair et al., 2021).

The $Q^2$ values (Table 3i) for Efficiency ($Q^2$ = 0.533), Service Delivery ($Q^2$ = 0.613), and Strategic Outcomes ($Q^2$ = 0.261) are all greater than zero, supporting the model's predictive relevance. The Root Mean Square Error (RMSE) and Mean Absolute Error (MAE) values were also provided as additional measures of predictive accuracy (Geisser, 1975; Stone, 1974).

Our analysis supports all four hypotheses, suggesting that Business Alignment positively influences both Efficiency and Service Delivery in the context of digital transformation and thus empirically validates our model and the value of business architecture to organizations pursuing transformation. Furthermore, Efficiency plays a significant role in enhancing Service Delivery, which in turn contributes to achieving Strategic Outcomes.

**Table 3f: Structural estimates (hypothesis testing)**

| Hypotheses | Beta | T-value | Decision | f square |
|---|---|---|---|---|
| H1: Business Alignment → Service Delivery | 0.565 | 7.455 | Supported | 0.434 |
| H2: Business Alignment → Efficiency | 0.735 | 17.256 | Supported | 1.177 |
| H3: Efficiency → Service Delivery | 0.302 | 4.001 | Supported | 0.124 |
| H4: Service Delivery → Strategic Outcomes | 0.588 | 7.407 | Supported | 0.529 |

Notes: critical t-values. *1.96 (P<0.05); **2.58 (P<0.01)

**Table 3g: R-squared ($R^2$) and Adjusted R-Squared (Adj. $R^2$) values**

| Construct | $R^2$ | Adjusted $R^2$ |
|---|---|---|
| Efficiency | 0.541 | 0.537 |
| Service Delivery | 0.662 | 0.657 |
| Strategic Outcomes | 0.346 | 0.341 |

**Table 3h: Effect sizes ($f^2$)**

| Construct | Business Alignment | Efficiency | Service Delivery | Strategic Outcomes |
|---|---|---|---|---|
| Business Alignment | | 1.177 | 0.434 | |
| Efficiency | | | 0.124 | |
| Service Delivery | | | | 0.529 |
| Strategic Outcomes | | | | |

**Table 3i: Predictive relevance $Q^2$ (PLSPredict)**

| Construct | Q2 Predict | RMSE | MAE |
|---|---|---|---|
| Efficiency | 0.533 | 0.695 | 0.543 |
| Service Delivery | 0.613 | 0.633 | 0.482 |
| Strategic Outcomes | 0.261 | 0.874 | 0.668 |

*Impacts of Business Architecture in the Context of Digital Transformation: An Empirical Study Using PLS-SEM Approach**4.3 Discussion*
This study aimed to explore the role of business architecture in the success of digital transformation initiatives. Based on the adapted balanced scorecard (BSC) method (Kaplan & Norton, 2005) and the business architecture impact model (AL-Malaise AL-Ghamdi, 2017), the study focused on four key dimensions of digital transformation: business alignment, efficiency, service delivery, and strategic outcomes. In this section, we discuss the key findings and their implications.

*4.4 The relationship between Business Alignment and Service Delivery*
RQ1 aimed to explore the influence of Business Alignment on Service Delivery in the context of digital transformation. Specifically H1 posited that Business Alignment would positively impact Service Delivery. The empirical results supported this hypothesis (Table 3f), indicating a positive and significant relationship between Business Alignment and Service Delivery ($\beta = 0.565$, $t = 7.455$, $P < 0.05$). This finding aligns with previous research suggesting that effective alignment between business and IT strategies enhances the quality of service delivery in organizations undergoing digital transformation (Pappas et al., 2018; Tarhini et al., 2017). The positive relationship between Business Alignment and Service Delivery can be understood by the fact that effective integration of IT and business strategies enables organizations to better understand and address customer needs, streamline processes, and improve the overall quality of products and services (Bharadwaj et al., 2013). Moreover, business architecture facilitates the design of flexible and scalable solutions, which allows organizations to adapt to changes in customer demands and market conditions, thereby enhancing service delivery (Black et al., 2017; Kappelman et al., 2014).

*4.5 The relationship between Business Alignment and Efficiency*
RQ2 sought to examine the impact of Business Alignment on Efficiency in organizations undergoing digital transformation and H2 proposed that Business Alignment would have a positive impact on Efficiency. The analysis revealed a positive and significant relationship between Business Alignment and Efficiency ($\beta = 0.735$, $t = 17.256$, $P < 0.05$), supporting H2 (Table 3f). This finding is consistent with previous studies that have emphasized the importance of aligning business and IT strategies to achieve higher levels of operational efficiency (Chan & Reich, 2007; Gerow et al., 2014). The positive relationship of Business Alignment on Efficiency can be attributed to several factors. First, alignment facilitates the integration of IT resources and capabilities with organizational processes, which can lead to cost and cycle time reductions (Henderson & Venkatraman, 1999; Tangi et al., 2020; Vial, 2019). Second, standardized IT platforms and architectures that are aligned with business strategies can simplify and streamline operations, thus improving efficiency (Kane, 2019; Weill & Woerner, 2017). Finally, alignment enhances communication and collaboration between IT and business units, enabling organizations to identify and address inefficiencies more effectively (Iivari & Huisman, 2007; Tallon & Pinsonneault, 2011).

*4.6 The role of Efficiency in enhancing Service Delivery*
RQ3 aimed to investigate the role of Efficiency in enhancing Service Delivery during digital transformation initiatives and H3 suggested that Efficiency would have a positive impact on Service Delivery. The empirical analysis supported H3 (Table 3f), showing a positive and significant relationship between Efficiency and Service Delivery ($\beta = 0.302$, $t = 4.001$, $P < 0.05$). This finding corroborates prior research highlighting the role of operational efficiency in improving service delivery during digital transformation (Berman, 2012; Li et al., 2018). This positive association between Efficiency and Service Delivery underscores the important role of processes in DT and demonstrates that efficient processes enable organizations to reduce costs and cycle times, leading to faster response times and better resource utilization (Legner et al., 2017; Ray et al., 2005). Moreover, efficiency gains can be reinvested in the improvement of customer-facing processes, enabling organizations to enhance the overall quality of their products and services (Mithas et al., 2013; Setia et al., 2013). Additionally, efficiency improvements can free up resources that can be allocated to customer service and support functions, which in turn can enhance service delivery (Melville et al., 2004; Naimi-Sadigh et al., 2022).

*4.7 The contribution of Service Delivery in achieving Strategic Outcomes*
RQ4 sought to examine how improved Service Delivery contributes to achieving Strategic Outcomes in the context of digital transformation and H4 asserted that Service Delivery would have a positive impact on Strategic Outcomes. The results supported H4 (Table 3f), revealing a positive and significant relationship between Service Delivery and Strategic Outcomes ($\beta = 0.588$, $t = 7.407$, $P < 0.05$). This finding is in line with previous studies that have demonstrated the importance of service delivery in achieving strategic outcomes, such as competitive advantage and market performance (AL-Malaise AL-Ghamdi, 2017; Davenport, 1993; Kohli & Grover, 2008). Such as positive relationship between Service Delivery and Strategic Outcomes can be attributed to several factors. First, improved service delivery can lead to higher customer satisfaction and loyalty, which in turn can translate into increased market share and profitability (Anderson et al., 1994; Volberda et al., 2011). Second, effective service delivery can help organizations differentiate themselves from competitors, thereby creating a sustainable competitive advantage (Porter, 2008; Smith et al., 2008). Third, the ability to deliver superior products and services can enhance an organization's reputation and brand value, which can further contribute to the achievement of strategic outcomes (Keller et al., 2021; Kotler & Kotler, 2009; Morgan et al., 2004).

**Page | 80**

## 5. Conclusion
This study provides valuable insights into the critical role of business architecture in driving the success of digital transformation initiatives. Through an empirical investigation of the relationships between business alignment, efficiency, service delivery, and strategic outcomes, the research has demonstrated that the adapted balanced scorecard framework, combined with the PLS-SEM approach, can effectively measure and assess the impact of business architecture on digital transformation. The findings have significant implications for both theory and practice, highlighting the importance of aligning business and IT strategies to improve efficiency and service delivery, which can ultimately contribute to the achievement of strategic outcomes.

A significant contribution of this study is the development and validation of a new model that addresses the discriminant validity issue encountered in the AL-Malaise AL-Ghamdi (2017) model. This new model provides clearer distinctions between the dimensions of business alignment, efficiency, service delivery, and strategic outcomes, and benefits from the established robustness of the Balanced Scorecard approach. Our analysis supports all four hypotheses, suggesting that Business Alignment positively influences both Efficiency and Service Delivery in the context of digital transformation and thus empirically validates our model and hypotheses. The findings from this study underscore the essential role of business architecture in driving digital transformation success and offer valuable insights for organizations seeking to enhance their digital transformation initiatives.

By emphasizing the practical importance of integrating business architecture into the digital transformation planning and execution process from the outset, organizations can better understand and address the complexities and challenges associated with digital transformation, ultimately leading to more effective and sustainable outcomes. Further, we demonstrate that by developing robust business architecture capabilities and fostering a culture that embraces and supports business architecture, organizations can be better able to navigate the complex and dynamic digital landscape.

We highlight the importance of business architecture as a strategic enabler of transformation, which we believe will only continue to grow as most organizations are still only at a low level of maturity in their business architecture practices. By recognizing the value of business architecture and investing in its development, organizations can better position themselves to achieve long-term, sustainable digital transformation success and thrive in an increasingly competitive and digitalized world. Finally, while our study has made significant contributions to our understanding of the role of business architecture in digital transformation success, there is still ample opportunity for future research in this area.

### *5.1 Limitations and Areas for Future Research*
Despite the valuable insights provided by this study, there are some limitations that should be acknowledged. First, the sample size of the study was relatively small, which may limit the generalizability of the findings. Future research should seek to replicate the study using larger and more diverse samples to validate the results and enhance the generalizability of the findings. Second, the study relied on self-reported data, which may introduce bias due to social desirability or recall inaccuracies. Future studies should consider using alternative data collection methods, such as interviews or observations, to triangulate the findings and increase the robustness of the results.

Third, the study focused on the impact of business architecture on digital transformation in terms of business alignment, efficiency, service delivery, and strategic outcomes. While these dimensions provide a comprehensive view of digital transformation, future research could explore additional aspects, such as employee engagement, organizational culture, or governance, to gain a more holistic understanding of the role of business architecture in digital transformation. Lastly, this study employed a cross-sectional design, which may limit the ability to establish causal relationships between business architecture and digital transformation outcomes. Future research should consider adopting a longitudinal design to track the evolution of business architecture and its impact on digital transformation over time. This approach could provide valuable insights into the dynamics of business architecture and digital transformation and help identify the factors that contribute to their success or failure.

**Funding**: This research received no external funding.
**Conflicts of Interest**: The authors declare no conflict of interest.
**ORCID iD** Dennis O'Higgins - 0000-0003-2138-6160
**Publisher's Note**: All claims expressed in this article are solely those of the authors and do not necessarily represent those of their affiliated organizations, or those of the publisher, the editors and the reviewers.